\documentclass[%
 prl,%
 amsmath,amssymb,
 reprint,%
twocolumn,
letterpaper,
]{revtex4}

\usepackage{graphicx}   % need for figures
\usepackage{amsmath,amssymb}
\usepackage{subfigure}
\usepackage[colorlinks=true,bookmarks=false,citecolor=blue,urlcolor=blue]{hyperref} %latex w/dvips

\newcommand{\pa}{\partial}
\newcommand{\eq}[1]{\begin{equation} #1 \end{equation}}
\newcommand{\eqa}[1]{\begin{align} #1 \end{align}}

\newcommand{\nl}{\nonumber \\}

\newcommand{\bse}{\begin{subequations}}
\newcommand{\ese}{\end{subequations}}

\begin{document}

\title{Theory for the Acoustic Raman Modes of Proteins}

\author{Timothy DeWolf and Reuven Gordon}
\affiliation{Department of Electrical and Computer Engineering, University of Victoria, British Columbia, Canada}
\email{rgordon@uvic.ca}

\date{\today}

\begin{abstract}
We present a theoretical analysis that associates the resonances of extraordinary acoustic Raman (EAR) spectroscopy [Wheaton et al., Nat Photon 9, 68 (2015)] with the collective modes of proteins.  The theory uses the anisotropic elastic network model to find the protein acoustic modes, and calculates Raman intensity by treating the protein as a polarizable ellipsoid.  Reasonable agreement is found between EAR spectra and our theory.  Protein acoustic modes have been extensively studied theoretically to assess the role they play in protein function; this result suggests EAR as a new experimental tool for studies of protein acoustic modes.  
\end{abstract}

\maketitle

%\section{Introduction}

The central dogma of molecular biology involves one-way information transfer from DNA to protein, a process that directly and reliably associates a particular three-dimensional structure with a given amino acid sequence.  Each structure is associated with a function (or functions); often, for example, a particular structure catalyzes a chemical reaction with remarkable selectivity.  The vibrational modes of a protein reflect its structure and conformation, and are thought to facilitate allostery and conformational change \cite{nmafunc, nmabook, Yang2007920, Dobbins29072008, coupl, Tama01012001}.  Of these modes, those with the lowest frequency are termed acoustic modes, and represent the largest thermal fluctuations of the protein.

Whereas many spectroscopic methods can probe localized resonances in a protein \cite{ref:proteinnmr,ref:prinInstrumentalAnalysisBook}, delocalized collective modes and their role in biological function have been historically difficult to measure \cite{thzOke}.  In the gigahertz (GHz) to low terahertz (THz) spectral window, electromagnetic absorption experiments have to deal with high solvent absorption and dielectric mixtures \cite{thzAbs1,thzAbs2,ghzdielecspec}.  Other experimental techniques for studying acoustic protein modes include inelastic incoherent neutron scattering (IINS) \cite{iinsprotein} and optical Kerr-effect (OKE) spectroscopy \cite{thzOke}.

\begin{figure*}[]
\centering
\subfigure[\label{fig:mode}\hspace{10pt}ANM eigenmode.]{\includegraphics[width=.21\textwidth]{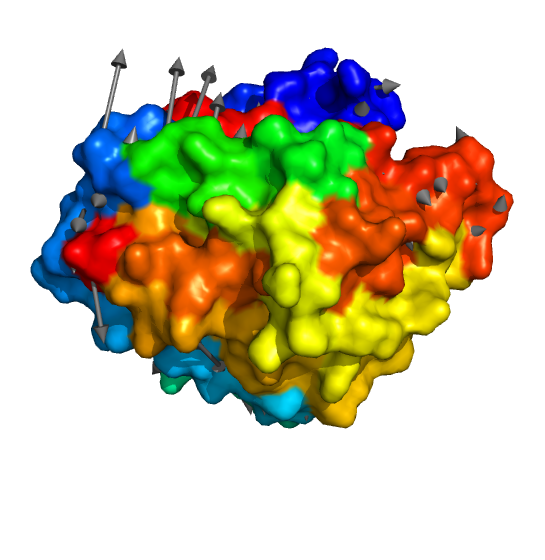}}
\includegraphics[width=.05\textwidth]{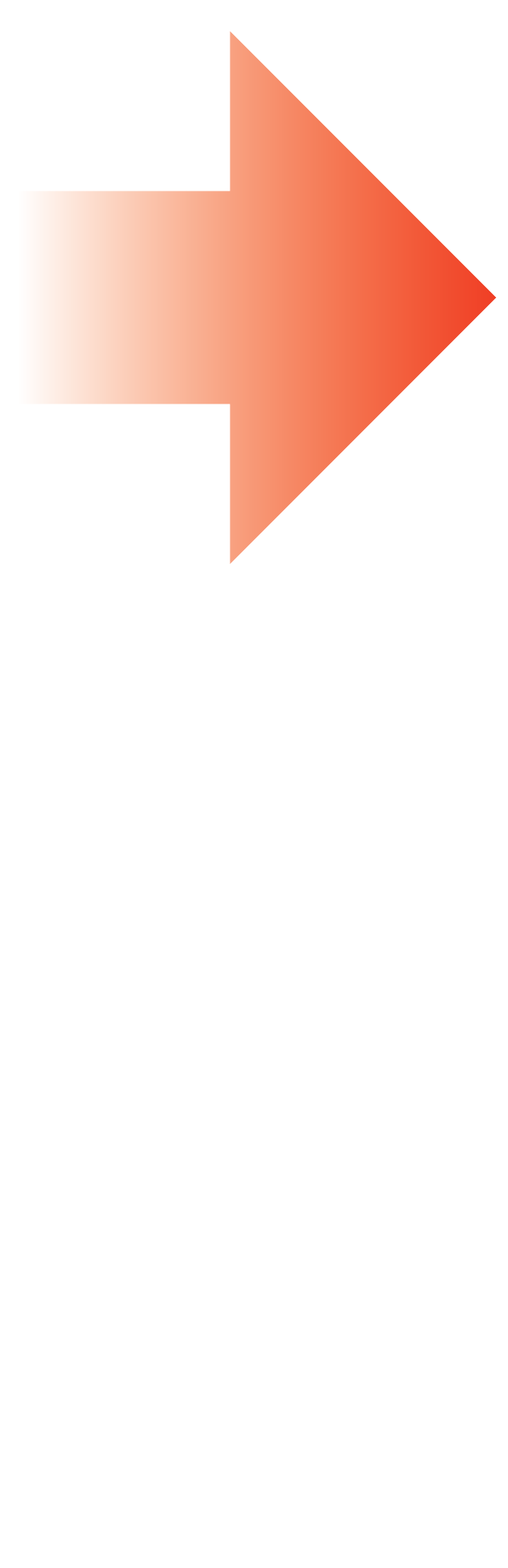}
\subfigure[\label{fig:eigvals}\hspace{10pt}Frequency spectrum, plotted with unit amplitudes.]{\includegraphics[width=.33\textwidth]{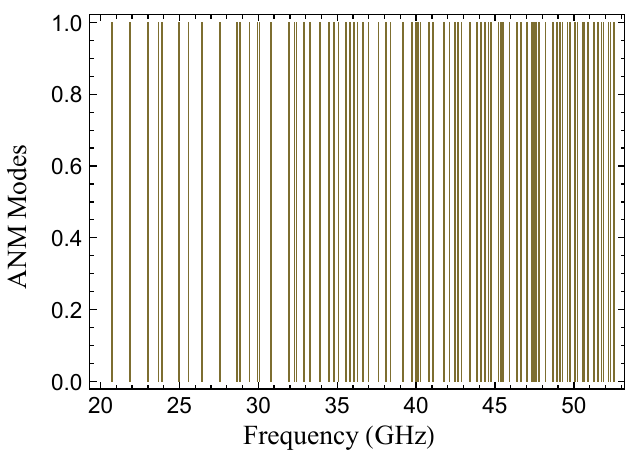}}
\includegraphics[width=.05\textwidth]{arrow.pdf}
\subfigure[\label{fig:ramansel}\hspace{10pt}Raman intensity spectrum (same protein modes as (b)).]{\includegraphics[width=.33\textwidth]{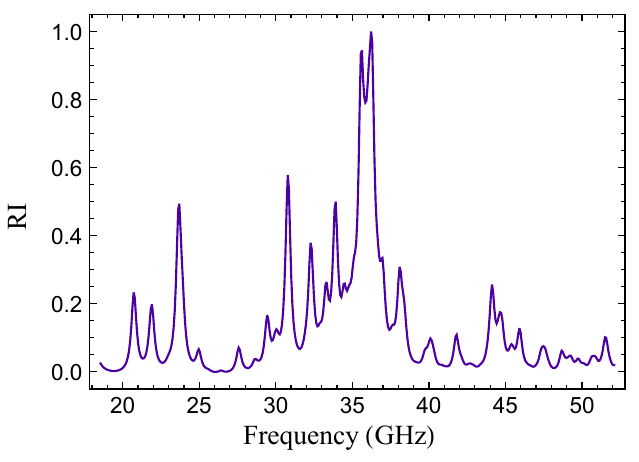}}
\caption{\label{fig:process} The method by which our protein Raman spectra are computed (shown here for carbonic anhydrase).  ANM yields a set of eigenvalues and eigenvectors; a sample eigenvector is shown in (a).  The set of eigenvalues (related to the frequencies) is shown in (b).  The eigenmodes are applied to generate positive and negative displacements for each mode, and the resulting protein shapes are fit to ellipsoids.  Analytic expressions are used to compute a Raman intensity for each mode.  The spectrum shown in (c) centers a Lorentzian function, multiplied by its Raman intensity (RI), at the position of each mode shown in (b).  As seen, a small subset of the original ANM modes are dominant in the Raman spectrum.}
\end{figure*}

We recently reported extraordinary acoustic Raman (EAR) spectroscopy as a way to measure resonances of optically trapped nanoparticles \cite{natsky}.  In EAR, the $\sim$10 to 100 GHz beating of two trapping lasers creates increased RMS fluctuation when the beat frequency matches a Raman-active particle resonance.  The frequency of vibrational modes in single polystyrene nanospheres were shown to fit with Lamb's theory.  The EAR spectra of several proteins were measured, however the remaining challenge is  ``...to associate the observed [protein] resonances with specific motions..." \cite{natreply}.  Here we propose a theory that assigns the measured EAR modes to low-frequency Raman-active protein modes.

Our theory uses elastic network model (ENM) normal mode analysis; elastic network models reproduce the essential dynamics of low-frequency protein modes to good accuracy \cite{PhysRevLett.77.1905}.  We use the ENM known as the anisotropic network model (ANM) \cite{anm2,Atilgan2001505}, as implemented in ProDy \cite{Bakan01062011}.

ANM represents the potential surface of an $N$ atom protein (excluding hydrogen) using a network of springs with spring constant $k$.  ANM analyses are often done using a reduced set of atoms, namely the $C_\alpha$ atoms along the protein backbone; we use an all-atom approach (excluding hydrogen) to build the elastic network.  Each spring connects a pair of atomic coordinates, but only atoms within cutoff radius $r_c$ are connected.  The matrix of second derivatives (taken with respect to the Cartesian coordinates of each atom) of this potential, known as the Hessian, is then computed; it is an $N\times N$ matrix of $3 \times 3$ (the protein coordinates are in $\mathbb{R}^3$) super-elements and has the units of $k$.  The diagonalization of this matrix yields $3N-6$ non-zero eigenvalues $\lambda_i$ and eigenvectors $Q_i$ that correspond to the frequency $\omega_i = \sqrt{ \lambda_i / m }$ and the displacement from equilibrium of each mode $i$.  $m$ is the mass of an atom; we use 13.2 amu for all atoms (a weighted average).  The six zero-valued eigenvalues correspond to rotational and translational degrees of freedom.  Fig.~\ref{fig:mode} shows what one of these eigenvectors $Q_i$ looks like for a protein.  At this point, we have a set of mode frequencies, as shown in Fig.~\ref{fig:eigvals}.

To calculate the intensity of each mode, we need to compute the Raman intensity of a given ANM mode from positive and negative coordinate displacements $\vec{r}_{i,\pm}=\vec{r}_0\pm \beta \vec{Q}_i$.  The equilibrium coordinates are $\vec{r}_0$; $\vec{Q}_i$ is a unit vector in $\mathbb{R}^{3N}$ and $\beta$ is a small scaling parameter.  We construct a quantity closely related to the inertia tensor, and diagonalize it to find the semi-principle axes (unit vectors) and lengths $a_{i,\pm},b_{i,\pm},c_{i,\pm}$ of two best-fitting ellipsoids \cite{0004-637X-548-1-68}, one for each of the stretched protein coordinates.  Dielectric polarizability tensors $\alpha_{i,\pm}$ for these best-fitting ellipsoids are then computed, using analytic expressions \cite{sihvola1999electromagnetic,doi:10.1080/14786444508521510} that require only the semi-principle axis lengths and the internal and external relative dielectric permittivity.  We take the two permittivity values to be $\epsilon_i=n^2=1.6^2$ (protein) \cite{MCMEEKIN1962151} and $\epsilon_e=1.33^2$ (water).

\begin{table*}[]
\centering
\begin{tabular}{lcccc}
\hline
\hline
\multicolumn{1}{p{2.5cm}}{\centering \textbf{Name} } & \multicolumn{1}{p{2.7cm}}{\centering \textbf{Molecular Weight (kDa)}} & \textbf{PDB ID} & \multicolumn{1}{p{4.1cm}}{\centering \textbf{ ANM Spring Constant \emph{k} (kJ mol$^-1$ \AA$^{-2}$)}} \\%& \multicolumn{1}{p{2.7cm}}{\centering \textbf{Fraction of Raman-Active Modes}}\\
\hline
pancreatic trypsin inhibitor &  6.6& 5PTI \cite{Wlodawer1984301} & 1.51\\
carbonic anhydrase I & 29.7 & 1CRM \cite{NYAS:NYAS49} &1.28\\
streptavidin & 52.8& 3RY2 \cite{3RY2} &1.29\\
ovotransferrin &  76.2& 1OVT \cite{Kurokawa1995196}&0.79\\
cyclooxygenase-2 & 274.4 & 5COX \cite{Kurokawa1995197}&1.15\\
\hline
\hline
\end{tabular}

\vspace{5pt}
\caption{\label{proteins} Summary of proteins measured experimentally and the associated PDB structure files (coordinate data) used here.}
\end{table*}

The Raman polarizability $\alpha'_i = (\pa \alpha_i / \pa Q_i)_0$ measures the change in polarizability due to the difference between the protein coordinates $\vec{r}_{i,\pm}$ \cite{1970raman}.  We calculate this as $\alpha'_i \sim \alpha_{i,+} -\alpha_{i,-}$.  We also account for the possibility that the two best-fit ellipsoids have rotated under the action of the pair of mode displacements by rotating one of the polarizability tensors as a rank two tensor, using the rotation matrix formed using the pair of best-fit semi-principle axes \cite{hand1998analytica}.  With this, the Raman intensity $I_i$ of ANM mode $i$ is given by \cite{1970raman}
\eq{ \label{ramani} I_i \sim 45 \bar{\alpha}'^2_i + 4\gamma'^2_i }
where
\eqa{ \bar{\alpha}_i' &= \frac{1}{3} (\alpha'_{i,xx} + \alpha'_{i,yy} + \alpha'_{i,zz} ) \\
\gamma'^2_i &= \frac{1}{2} ( (\alpha'_{i,xx} - \alpha'_{i,yy} )^2 + (\alpha'_{i,yy} - \alpha'_{i,zz} )^2\nl
& + (\alpha'_{i,zz} - \alpha'_{i,xx} )^2 +6(\alpha'^2_{i,xy}+\alpha'^2_{i,yz}+\alpha'^2_{i,zx} ) ). }
%\gamma'^2 &= \frac{1}{2} ( (\alpha'_{xx} - \alpha'_{yy} )^2 + (\alpha'_{yy} - \alpha'_{zz} )^2 + \nl
%&+(\alpha'_{zz} - \alpha'_{xx} )^2+6(\alpha'^2_{xy}+\alpha'^2_{yz}+\alpha'^2_{zx} ) ). }
The mean value $\bar{\alpha}'_i$ measures change in polarizability of a particular mode due to linear stretching; $\gamma'_i$ gives the anisotropic contribution.  Spectra (see for example Fig.~\ref{fig:ramansel}) are constructed by centering Lorentzian functions at the frequency position $\omega_i$ of each mode, with mode heights proportional to the Raman intensities $I_i$ (and plotting the summation of these curves).  We choose a constant Lorentzian linewidth for each spectrum. % The collective nature of a mode and its Raman intensity are independent; the lower frequency modes generally feature more collective motion than the high frequency modes (see Fig.~4 in the supporting information), whether or not a mode has high Raman activity.

We compare our theory with previously published experimental data \cite{natsky} for the five proteins listed in Table~\ref{proteins}, and list the Research Collaboratory for Structural Bioinformatics Protein Data Bank (RCSB PDB) \cite{Berman01012007} structures used in computation.  (The streptavidin data was not published but was acquired during the same period.)  It is assumed that the PDB crystal coordinates are close to the potential minimum (i.e. that the crystal coordinates approximately give $\vec{r}_0$).  By matching the atomic mean-square fluctuations predicted by ANM (calculated using ProDy) with the crystallographic isotropic temperature factors included with the PDB crystal data, we associate a $k$ with each protein \cite{Atilgan2001505}.  These $k$ values are shown in Table~\ref{proteins}.  Details of these ANM, spring constant, ellipsoid fitting and Raman calculations are given in the supporting information.

\begin{figure*}[]
\centering
\textbf{ \hspace{30pt} EXPERIMENT \hspace{160pt} THEORY }
\includegraphics[width=.93\textwidth]{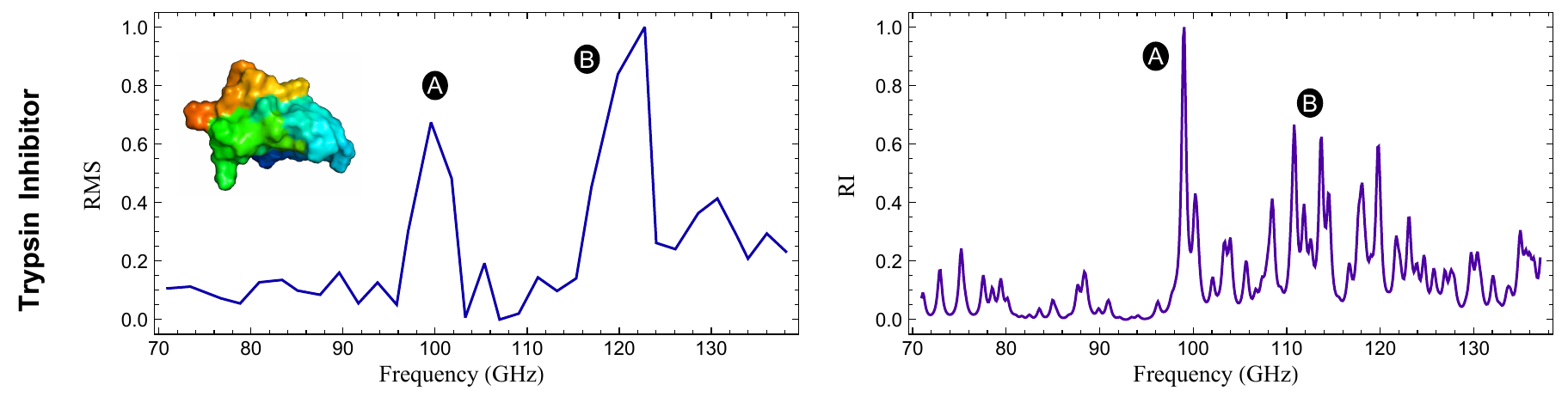}
\includegraphics[width=.93\textwidth]{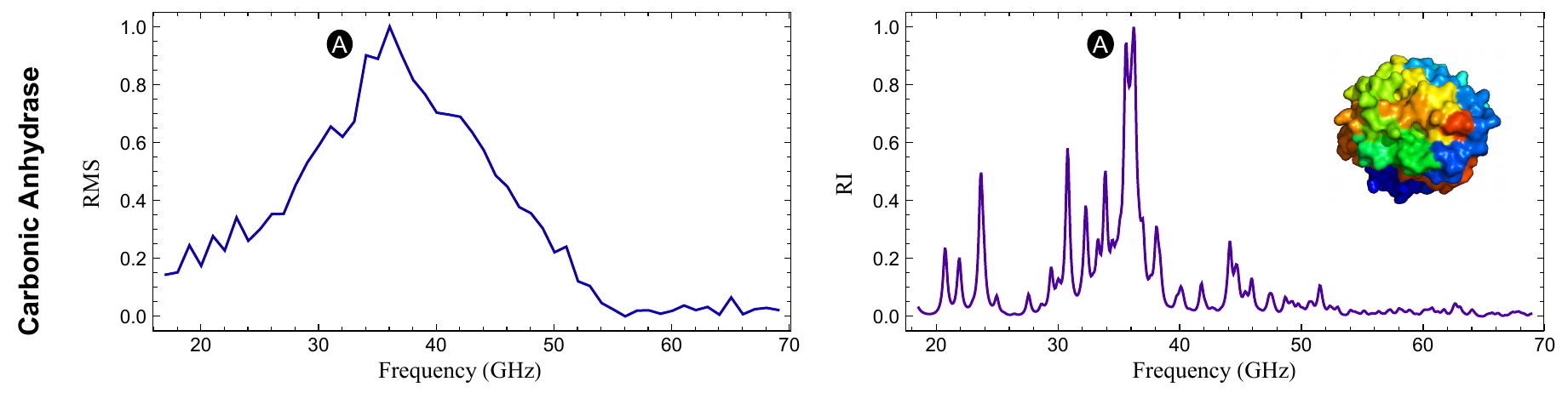}
\includegraphics[width=.93\textwidth]{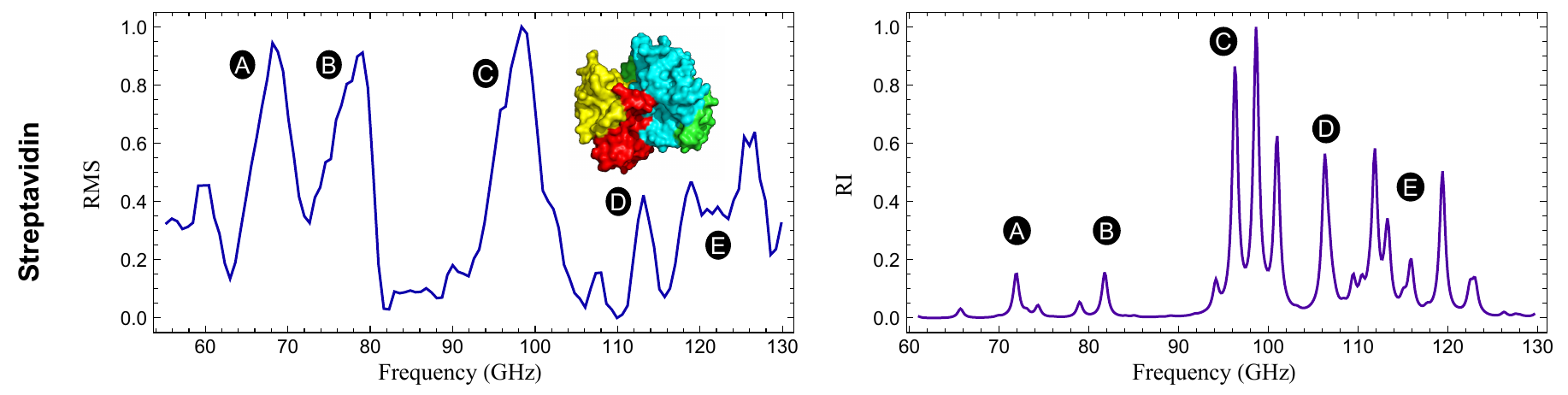}
\includegraphics[width=.93\textwidth]{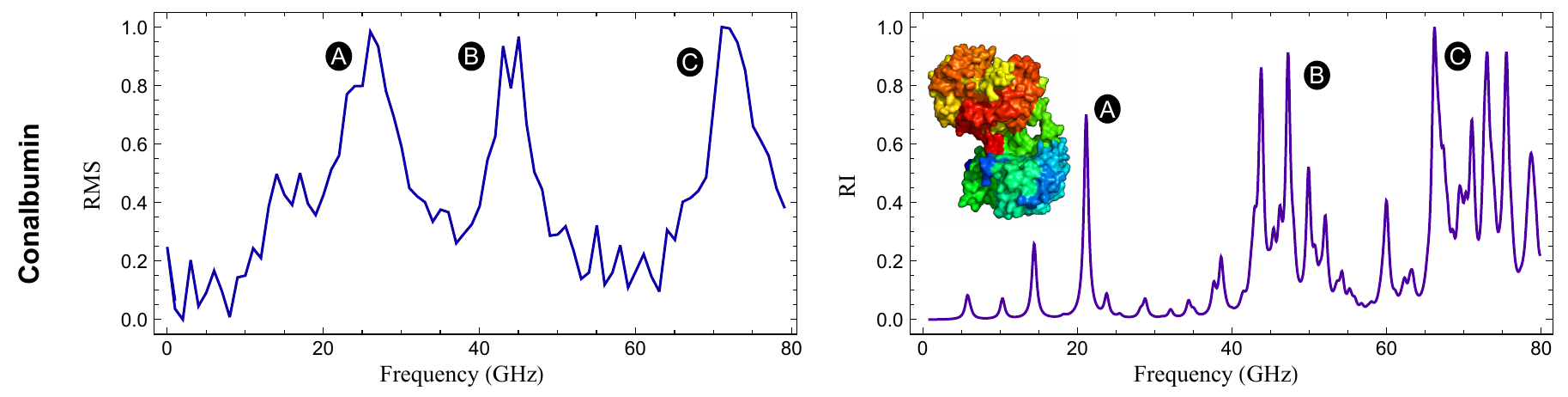}
\includegraphics[width=.93\textwidth]{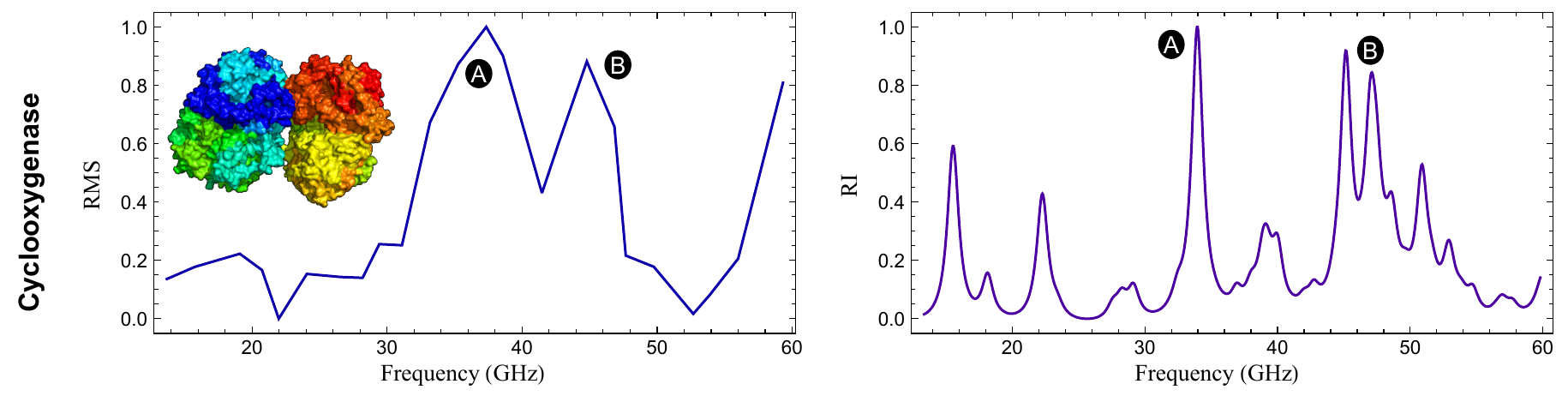}
\caption{\label{fig:theoryexpPlots}A comparison of the experimentally measured extraordinary acoustic Raman (EAR) spectra (left) and elastic network/Raman ellipsoid polarizability model spectra (right).  RMS is the root-mean-squared variation in the optical trap transmission.  RI is the theoretical Raman intensity, calculated using an all-atom ANM with cutoff radius of $r_c$=7.9 \AA.  Suggested correlations between theory and experiment of selected peaks (or groups of peaks) are letter-labelled.}
\end{figure*}

The ANM cutoff distance $r_c=7.9$ \AA$ $ was selected by hand for best overall agreement between theory and experiment for the five proteins.  At values near this $r_c$, EAR mode frequencies $\tilde{\omega}_i$ and ANM frequencies $\omega_i$ are approximately linearly proportional: $\tilde{\omega}_i = \zeta \omega_i$.  The proportionality constant $\zeta$ is a free parameter in our theory; a $\zeta$ is chosen for each protein (see supporting information).  It is the fine spectral resolution of EAR that allows us to directly fit the spectral modes for each protein; in prior works obtaining a Gaussian-distributed density of states has been used as a criterion for selecting physical values for $r_c$ \cite{Atilgan2001505}.  

%\section{Discussion}

The computed spectra along with our previously obtained experimental EAR spectra are shown in Fig.~\ref{fig:theoryexpPlots}.  The visual agreement between theory (right side) and experiment (on the left) in Fig.~\ref{fig:theoryexpPlots} is, in our opinion, quite remarkable.  The intensity and frequency placement of the major peaks, as well as some of the minor peaks, agree with the experimental data.  We have made many approximations, including the use of a ``spring network" potential in place of a more realistic potential map (e.g. the semiempirical potentials employed in molecular dynamics) and the representation of protein polarizability by the polarizability of a dielectric ellipsoid.  The normal modes could also be computed in the time domain, by combining molecular dynamics simulation with principal component analysis to accurately capture the low-frequency modes \cite{mdstructurefunction}. Wider bandwidth Raman intensity spectra for each protein are given in the supporting information.  As can be seen in the extended spectra, the EAR data shown here have captured most of the major Raman-active collective modes in these five proteins.

In the light of our theory, we make some comments regarding the experiment.  A past optical trapping work by our group \cite{doi:10.1021/nl203719v} reported on a special type of conformational change, the N-F transition, found in bovine serum albumin (BSA); this conformation change can be viewed as a type of denaturation as it involves a reversible unfolding of BSA domain III \cite{rosenoer2014albumin, bsad}.  BSA can also be irreversibly denatured \cite{FEBS:FEBS469}.  We assume here, as we did in the EAR experiment \cite{natsky}, that the trapped proteins have not been irreversibly denatured.  We also assume that the proteins are not being reversibly unfolded or deformed by the optical forces, so that the equilibrium coordinates given by the x-ray crystal data will be good approximations of the optically trapped protein coordinates.  Past works (e.g. \cite{anha1,anha2}) have suggested that anharmonic effects play a role in the low frequency modes of proteins, whereas our ANM-based theory is a purely harmonic model of protein motion.  In a Duffing oscillator, for example, the absence of nonlinear effects such as jumping, hysteresis, and bistability can be related to the fact that harmonic driving force is sufficiently weak \cite[Chapter~7]{enns2012nonlinear}.  Thus an explanation for the seeming unimportance of anharmonicity in our theory is that the amplitude of the driving force is low enough that the protein response is linear.

There has hitherto been relatively little experimental evidence for the existence of protein collective modes---accomplishments in protein THz spectroscopy have not been able to conclusively connect measurements with biologically relevant collective protein motions \cite{thzOke}.  There has also been difficulty in assigning physical frequency units to ENM.  These results directly connect ENM mode analysis to EAR.  They provide another way to validate ENM results, and suggest EAR as a new tool for future experimental studies of low-frequency protein collective modes.  They also suggest that EAR may provide a way to improve ENMs.

We would like to acknowledge the use of the computational resources of WestGrid (www.westgrid.ca) and Compute Canada (www.computecanada.ca).  This work was supported in part by an NSERC Discovery Grant and funding from the Faculty of Graduate Studies at the University of Victoria.  The protein renderings were prepared using PyMOL \cite{PyMOL} and POV-Ray \cite{pov}.

\bibliography{nmaPaper}

\end{document}